\documentclass[showpacs,prb,aps,onecolumn,epsf,amsmath,amssymb]{revtex4}

\usepackage[xdvi]{graphicx}
\usepackage{feynmp}

\newcommand{\BF}[1]{\mbox{\boldmath $#1$}}
\newcommand{\BFS}[1]{\mbox{\scriptsize\boldmath $#1$}}

\begin{document}
\title{Intra-Landau level polarization effect for a striped Hall gas} 
\author{
T. Aoyama
\footnote{
Present address: 
Institute Henri Poincar\'e, 11 rue Pierre et Marie Curie,
75231 Paris cedex 05 France, 
},
K. Ishikawa,
Y. Ishizuka and N. Maeda}
\address{
Department of Physics, Hokkaido University, 
Sapporo 060-0810, Japan
}
\date{\today}
\begin{abstract}
We calculate the polarization function 
including only intra-Landau level correlation effects of striped Hall gas. 
Using the polarization function, 
the dielectric function, the dispersion of the plasmon and the correlation 
energy are computed in a random phase approximation (RPA) and 
generalized random phase approximation (GRPA). 
The plasmon becomes anisotropic and gapless owing to the anisotropy 
of the striped Hall gas and two dimensionality of the quantum Hall system. 
The plasmon approximately agrees with the phonon derived before 
by the single mode  approximation. 
The (G)RPA correlation energy is compared with other numerical calculations. 
\end{abstract}
\draft
\maketitle
\section{Introduction}
At the half-filled third and higher Landau level (LL), 
the anisotropic longitudinal resistance 
has been observed in ultrahigh mobility samples.\cite{Lilly,Du} 
Collective excitations and related physical quantities of the 
anisotropic state give important information to clarify this state, 
but have not been studied well experimentally and theoretically. 
In this paper, we study collective excitations by calculating a polarization
function in intra-LL. 

The Hartree-Fock approximation (HFA) at a half-filled $l$th LL in
$l\rightarrow \infty$ limit 
predicts that a unidirectional charge density wave (UCDW) forms 
in the two-dimensional (2D) electron system.\cite{Shk,Moes} 
Within the intra-LL, 
the HFA at the half-filled $l$th LL ($l\geq 2$) shows 
that an anisotropic Fermi surface forms in the UCDW which explains 
the anisotropic longitudinal resistance and exotic 
response.\cite{Ma,opt,review} 
Collective modes for the UCDW are studied based on an edge current picture
\cite{Anna} and a single mode approximation (SMA).\cite{SMA}   
We call the UCDW the striped Hall gas. 
Another solution in the HFA is a modulated striped state 
which is highly anisotropic charge density wave (ACDW) having an energy
gap and no Fermi surface. 
Collective modes for ACDW is derived by the time dependent 
HFA (TDHFA).\cite{Cote} 
In this paper, we study the striped Hall gas by using a random phase 
approximation (RPA), in which bubble diagrams are summed up, and a 
generalized RPA (GRPA), in which bubble and ladder diagrams are summed up. 
The spectrum of collective modes and correlation energy of the striped Hall 
gas are compared with the results of SMA  and ACDW. 

The electron gas system in the absence of a magnetic field 
has been studied  successfully by using diagram techniques 
in many-particle physics. 
Diagram technique enables us to compute corrections to 
HFA systematically.\cite{gellmann} 
In a 2D electron system in a strong magnetic field, on the other hand, 
the electron kinetic energy is frozen to one LL and it seems difficult to 
analyze the system by using diagram techniques. 
In the von Neumann lattice (vNL) formalism,\cite{E} however, 
it is possible to apply a systematic diagram technique to 
the 2D electron system under a magnetic field.
In the vNL formalism, the 2D electron system under a homogeneous magnetic 
field is 
represented as a 2D lattice system and a momenta is defined in the 
magnetic Brillouin zone (MBZ). 
The electron kinetic energy is induced in HFA within intra-LL and depends on 
the momentum. 
Hence the free propagator is defined by using the kinetic energy and a diagram 
technique is applied systematically. 

In this paper,  we investigate quantum fluctuation effects of the striped 
Hall gas below the cyclotron energy scale. 
From the  polarization function including only intra-LL effects 
at one-loop order, 
we obtain the dielectric function, the plasmon and 
the correlation energy in the (G)RPA. 
Isotropic screening effects in the dielectric function at higher LL 
have been estimated.\cite{Aleiner} 
The intra-LL screening effects in the dielectric function is highly 
anisotropic for the striped Hall gas. 
The plasma frequency which is obtained by the zero of the dielectric function 
in (G)RPA, is found to be an anisotropic gapless mode. 
In the RPA, the energy of plasmon is larger than the particle-hole
excitation as a usual plasmon in the electron gas without a magnetic field. 
In the GRPA, on the other hand, the energy of the plasmon becomes smaller than 
the particle-hole excitation. 
The plasmon in the GRPA approaches to the phonon derived by the SMA as 
including a long-range component of the Coulomb interaction. 
The correlation energy in the (G)RPA substantially reduces the total energy. 

This paper is organized as follows. 
The striped Hall gas with the Fermi surface of a strip shape is 
introduced in Sec.~\ref{II}. 
In Sec.~\ref{Intra}, we calculate the polarization function at the one-loop 
order and derive the dielectric function, the plasmon 
and the correlation energy in the (G)RPA. 
Summary is given in Sec.~\ref{IV}. 
Appendix~\ref{HF} gives the LL-projected HF approximation in the vNL 
formalism. 
The explicit one-electron energy of the striped state is given in 
Appendix~\ref{appe_a}.  
In Appendix~\ref{Feynman_rule}, the Feynman rule for the 
diagram techniques are presented. 
In Appendix~\ref{2-loop}, a duality relation between the direct term and the 
exchange term is shown. 

\section{Striped Hall gas with an anisotropic Fermi surface}
\label{II}
In this section, 
we derive the striped Hall gas at the half-filled higher LL in the
vNL formalism.\cite{Imo,Ma} 
In the vNL formalism, 
the 2D electron system in a perpendicular uniform
magnetic field is transformed into a 2D lattice system and 
the momentum becomes good quantum number of the 
one-electron state. 
In the HFA, the kinetic energy is induced by the interaction term 
and perturbative calculation is easily carried out by using the 
diagram technique. 

Let us consider a 2D electron system in a perpendicular 
uniform magnetic field $B$.
The total Hamiltonian $H$ of the system is written as $H=H_0+H_1$, 
\begin{eqnarray}
H_0&=&\int \psi^\dagger({\bf r},t){({\hat{\bf p}}+e{\bf A})^2\over2m}
\psi({\bf r},t)d^2r,\nonumber\\
H_1&=&{1\over2}\int
:\rho({\bf r},t)V({\bf r}-{\bf r}')\rho({\bf r}',t): d^2r
d^2r',\label{Eq:kch}
\end{eqnarray}
where ${\hat p}_\alpha=-i\hbar\partial_\alpha$, 
$\partial_x A_y-\partial_y A_x=B$, 
$V=q^2/\vert\bf r\vert$, $q^2=e^2/4\pi\epsilon$
($\epsilon$ is a background dielectric constant).
$\psi({\bf r},t)$ is the electron field, 
and $\rho({\bf r},t)=\psi^\dagger({\bf r},t)\psi({\bf r},t)$. 
The symbol $:$ means the normal ordering for the vacuum of one electron 
field. 
In the following calculation, the units are set as $\hbar=c=1$. 
$H_0$ is the free Hamiltonian, which is quenched in a LL. 
$H_1$ is the Coulomb interaction. 
Since we want to consider only intra-LL effects, the free Hamiltonian is
omitted. 
We ignore the spin degree of freedom.
The electron field is expanded by the momentum state $\vert f_l
\otimes\beta_{\bf p}\rangle$ in the vNL formalism\cite{E} as 
\begin{equation}
\psi({\bf r},t)
=
\int_{\rm MBZ} {d^2 p\over(2\pi)^2}\sum_{l=0}^\infty b_{l}
({\bf p},t)
\langle{\bf r}\vert f_l\otimes\beta_{\bf p}\rangle, 
\end{equation}
where $b_{l}({\bf p},t)$ is the anti-commuting annihilation operator
with momentum $\bf p$ defined in the MBZ. 
The base function depends on two momenta 
$p_x$, $p_y$ symmetrically. 
The annihilation operator $b_{l}({\bf p},t)$ obeys a twisted periodic boundary
condition $b_{l}({\bf p}-2 \pi {\bf n},t)
=e^{-i \pi (n_x+n_y)+i n_y p_x} b_{l}({\bf p},t)$, 
where $n_x$, $n_y$ are integers.
The momentum state is the Fourier transform of the Wannier basis of 
vNL which localized at ${\bf r}=a (r_s m,n/r_s)$, 
where $n$, $m$ are integers. 
Here $a=\sqrt{2\pi\hbar/eB}$, and $r_s$ is a vNL asymmetry parameter. 
In this paper, 
we set $a$ to be $1$. 
The MBZ means $|p_\alpha|\leq \pi$.
The $t$ dependence of the operators is omitted for simple notation. 
The Fourier transformed current operator $j^\mu({\bf k})$ 
is written in the vNL formalism as 
\begin{eqnarray}
j^\mu({\bf k})
=
\int_{\rm MBZ}{d^2p\over (2\pi)^2} \sum_{ll'} b_{l}^\dagger({\bf p})
b_{l'}({\bf p}-{\hat{\bf k}})
\langle f_l \vert {1\over 2}
\{ v^\mu,e^{-i{\bf k}\cdot\BFS{\xi}}\} \vert f_{l^\prime} \rangle
 \exp\left[-i{1\over4\pi}{\hat k}_x(2p_y-{\hat k}_y)\right],
\label{Eq:rho}
\end{eqnarray}
where ${\hat{\bf k}}=(r_s k_x,k_y/r_s)$ and $\BF{\xi}$ is 
the relative coordinate of the electron and $v^\mu=[1,-(eB/m)\eta,(eB/m)\xi]$, 
$\mu=$0, 1, 2. 
In order to project the operators into the $l$th LL, we take only 
$l$th LL index in Eq.~(\ref{Eq:rho}) and write it as $j^\mu_\ast({\bf k})$. 
We define $g_l^\mu(k)=\langle f_l \vert {1\over 2}
\{ v^\mu,e^{-i{\bf k}\cdot\BFS{\xi}}\} \vert f_l \rangle$. 
For the projected density operator $\tilde{\rho}_\ast({\bf k})=
j^0_\ast({\bf k})$, 
we use the short-hand notation $f_l(k)=g_l^0(k)$ and $f_l(k)=L_l({k^2/4\pi})
e^{-{k^2/8\pi}}$ where $L_l$ is the Laguerre polynomial. 
In the LL projected space, the Fourier transformed Coulomb interaction is 
modified 
as $V_l(k)=[f_l(k)]^2 2\pi q^2/k$ for the vNL operator $b_l({\bf p})$. 
We apply the HFA to the Coulomb interaction within the $l$th LL, 
and get the one-electron spectrum satisfying a self-consistency
explicitly (see Appendix~\ref{HF}). 

There are two conserved charges $Q_X$ and $Q_Y$ corresponding to the 
magnetic translation in $-y$ direction and  $x$ direction, respectively, 
\begin{eqnarray}
Q_X&=&\left.i\frac{\partial\tilde{\rho}_*({\bf k})}{\partial k_x}\right
\vert_{\bf{k}=0},
\label{eq:Q}\\
Q_Y&=&\left.i\frac{\partial\tilde{\rho}_*({\bf k})}{\partial k_y}\right
\vert_{\bf{k}=0},
\nonumber
\end{eqnarray}
which satisfy $[H,Q_X]=[H,Q_Y]=0$.\cite{SMA}  
Note that $Q_X$ and $Q_Y$ correspond to the magnetic translation in $p_x$ 
direction and $p_y$ direction in the momentum space, respectively. 
We give a self-consistent mean field solution which is uniform in the 
$y$ direction and periodic in the $x$ direction, that is the striped Hall gas. 
This state is given as
\begin{equation}
\vert{\rm HF}\rangle
=
N_1
\!\!\!\!\!\!\!\!\!\!\!\!
\prod_{\vert p_x\vert\le\pi,\vert p_y\vert\le\pi/2} 
\!\!\!\!\!\!\!\!\!\!\!\!
b_{l}^\dagger({\bf p})\vert 0\rangle,
\label{eq:fermi}
\end{equation}
where $\vert0\rangle$ is the vacuum for $b_{l}$ and $N_1$ is a 
normalization factor. 
This striped state 
satisfies the self-consistency Eq.~(\ref{eq:self}) 
at the half-filled higher LL.\cite{Imo,Ma} 
The corresponding one-electron energy 
has the anisotropic Fermi surface which is parallel to the $p_x$
axis. 
The explicit form of $\epsilon_l^{\rm HF}(p_y)$ is given by 
Eq.~(\ref{eq:spectrum}) in Appendix~\ref{appe_a}.
The density of this state $\langle {\rm HF} \vert\rho({\bf
r})\vert {\rm HF} \rangle$ is uniform in $y$ direction and periodic 
in $x$ direction with a period $r_s$.\cite{Ma,Imo} 
Using Eq.~(\ref{eq:fermi}), we can show 
\begin{eqnarray}
\langle{\rm HF}\vert[Q_X,\tilde{\rho}_*(\bf{k})]\vert{\rm HF}\rangle&=&0,\\
\langle{\rm HF}\vert[Q_Y,\tilde{\rho}_*(\bf{k})]\vert{\rm HF}\rangle&\neq& 0 
{\rm\ for\ }(k_x,k_y)=(2\pi n/r_s,0),
\nonumber
\end{eqnarray}
where $n$ is integer. 
Therefore the magnetic translational symmetry in $x$ direction or $p_y$ 
direction is spontaneously broken. 

The one-electron energy has an anisotropic energy gap in $p_x$
direction. 
The Fermi velocity, then, is in the $y$ direction in coordinate space. 
The orthogonality of the Fermi surface in the momentum space and the
density profile in the coordinate space is reminiscent of the Hall effect. 
The Hartree-Fock (HF) energy per particle is calculated as
$E_l^{\rm HF}(r_s)=\langle {\rm HF} \vert H_1\vert {\rm HF} \rangle/N$.  
We determine the optimal value $r_s=r_s^{\rm min}$ which corresponds to
the stripe period by minimizing $E_l^{\rm HF}(r_s)$.
At the half-filled $l=2$ LL, 
the optimal value is $r_s^{\rm min}=2.474.$\cite{Ma} 

\section{Intra-LL Polarization Effect}
\label{Intra}

In the $l$th LL Hilbert space, the free Hamiltonian $H_0$ is quenched and 
only the interaction Hamiltonian $H_1$ remains. 
Since there exists no bare kinetic term, it seems difficult to deal with the 
Coulomb interaction as a perturbative term, naively. 
In the HFA, however, the kinetic term of electrons is induced by 
the effective direct and exchange interaction. 
Hence, the effective kinetic term appears in the HFA and 
the quantum fluctuation around the HF ground state is caused by the 
residual interaction term. 
The Coulomb interaction $H_1$ is divided into two terms 
\begin{eqnarray}
H_1
=
H^{\rm HF}
+(H_1-H^{\rm HF})
=
H^{\rm HF}+
\frac{1}{2}\int_{-\infty}^{\infty} 
\!\!\!
\frac{d^2 k}{(2\pi)^2} 
\;_\circ^\circ \;
\tilde{\rho_\ast}({\bf k})
V_l(k)
\tilde{\rho_\ast}(-{\bf k})
\;_\circ^\circ \;. 
\end{eqnarray}
Here, 
$H^{\rm HF}$ is given by Eq.~(\ref{eq:HF}) in Appendix~\ref{HF}. 
Operators between the symbol $\:_\circ^\circ$ 
are normal ordered for the HF vacuum. 
We study the quantum fluctuation for the HF state using 
the polarization function in the (G)RPA. 
In the vNL formalism, 
we use the Feynman diagram technique in the vNL formalism which is 
presented in Appendix~\ref{Feynman_rule}. 
The dielectric function and excitation spectra of the striped Hall gas 
are given by the polarization function in the (G)RPA. 

\subsection{One-loop polarization function}
\begin{figure}
 \centerline{\includegraphics[width=.4\linewidth]{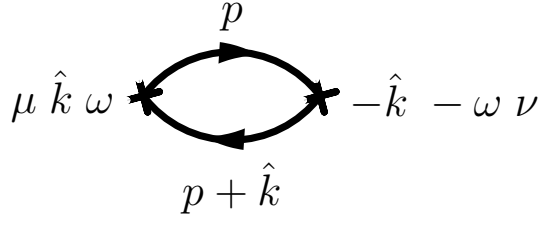}}
\caption{Feynman diagram for the current-current correlation function. }
\label{fig:polari}
\end{figure}

\begin{figure}[h]
\centerline{
 \includegraphics[width=.7\linewidth]{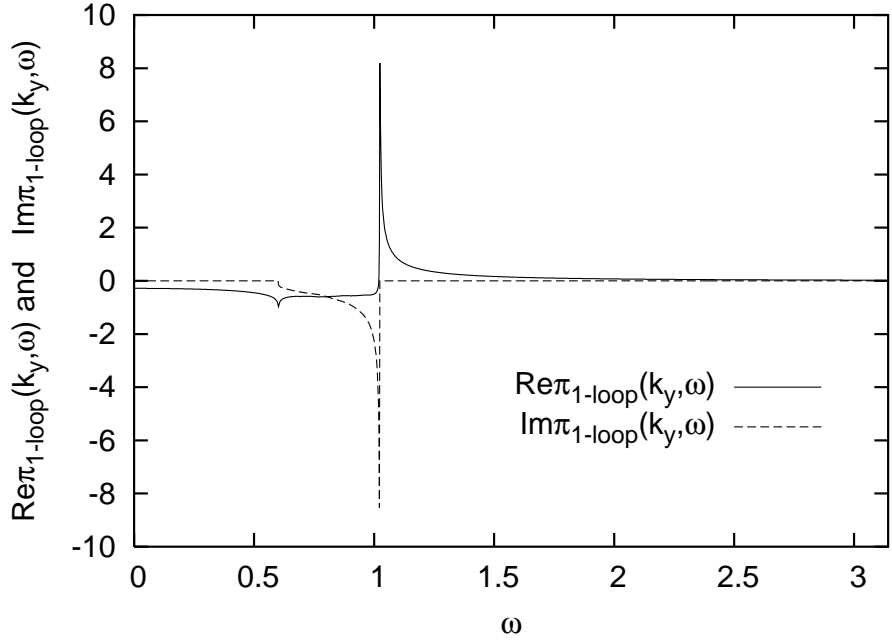}
}
\caption{$\omega$-dependence of the polarization function 
$\pi_{\rm 1-loop}(k_y,\omega)$ at $k_y=\pi/4$. 
The unit of $k_y$ is $r_s/a$. 
${\rm Im}\pi_{\rm 1-loop}$ vanishes at small $\omega$ region due to the
 strip shape of the Fermi surface. 
The unit of $\omega$ is $q^2/(a\hbar)$.}
\label{Pi}
\end{figure}

First let us study the polarization function in the one-loop order. 
The current-current correlation function, which is a response of 
the external electromagnetic field, in one LL is defined in the Heisenberg 
picture as
\begin{eqnarray}
\pi^{\mu\nu}({\bf k},\omega)
&=&
-i (TS)^{-1}\int_{-\infty}^{\infty} dt_1 dt_2
\langle \Psi_0 | {\rm T} \delta 
j^{\mu }_*({\bf k},t_1) \delta j^{\nu }_*(-{\bf k},t_2) 
|\Psi_0 \rangle e^{-i\omega (t_1-t_2)},
\label{Eq:pola_def}
\end{eqnarray}
where $\delta j^\mu=j^\mu-\langle \Psi_0 | j^\mu|\Psi_0\rangle$  and  $TS$ is 
a total time times a total area of a 2D electron system. 
The one-loop current-current correlation function 
shown in Fig.~\ref{fig:polari} is calculated as
\begin{eqnarray}
\pi_{\rm 1-loop}^{\mu\nu}({\bf k},\omega)
&=&
-i
g_l^\mu(k) g_l^\nu(k)
\int_{-\infty}^{\infty} \! \! \frac{d\omega_1}{2\pi}
 \int_{\rm MBZ} \!\! \frac{d^2p}{(2\pi)^2}
\;
\tilde{G}^{(0)}_{{\bf p},\omega_1}
\tilde{G}^{(0)}_{{\bf p}+\hat{\bf k},\omega_1+\omega}
\nonumber \\
&=&
g_l^\mu(k) g_l^\nu(k)\int_{\rm MBZ} \frac{d^2p}{(2\pi)^2}
\left[
\frac{\theta(\epsilon_{\bf p+\hat{k}}-\mu_{\rm F})
\theta(\mu_{\rm F}-\epsilon_{\bf p})}
{\omega-\epsilon_{\bf p+\hat{k}}+\epsilon_{\bf p}+i\delta}
-
\frac{\theta(\mu_{\rm F}-\epsilon_{\bf p+\hat{k}})
\theta(\epsilon_{\bf p}-\mu_{\rm F})}
{\omega-\epsilon_{\bf p+\hat{k}}+\epsilon_{\bf p}-i\delta}
\right]. 
\nonumber
\\
& &
\label{eqn:pi_k}
\end{eqnarray}
$\tilde{G}^{(0)}_{{\bf p},\omega}$ is the free electron propagator
given by Eq.~(\ref{eq:free_proper}). 
$\mu_{\rm F}$ is the Fermi energy. 
We define the one-loop polarization function in the $l$th LL by 
$\pi_{\rm 1-loop}({\bf k},\omega)=\pi^{00}({\bf k},\omega)/(g_l^0(k))^2$. 
On the basis that $\epsilon_{\bf k}$ is a even function, 
we can show that $\pi_{\rm 1-loop}(-{\bf k},\omega)
=\pi_{\rm 1-loop}({\bf k},\omega)$
and
$\pi_{\rm 1-loop}({\bf k},-\omega)
=\pi_{\rm 1-loop}({\bf k},\omega)$. 
The explicit form of the 
real and imaginary part of the one-loop polarization function are as follows: 
\begin{eqnarray}
{\rm Re}\pi_{\rm 1-loop}(k_y,\omega)
&=&
2 \lim_{\delta \to 0}
\int_{\frac{\pi}{2}-\hat{k}_y}^{\frac{\pi}{2}-\frac{\hat{k}_y}{2}}
\!\!
\frac{d p_y}{2\pi} 
\left[
\frac{\epsilon_{p_y}-\epsilon_{p_y+\hat{k}_y}-\omega}{
(\epsilon_{p_y}-\epsilon_{p_y+\hat{k}_y}-\omega)^2+\delta^2
}
+
\frac{\epsilon_{p_y}-\epsilon_{p_y+\hat{k}_y}+\omega}{
(\epsilon_{p_y}-\epsilon_{p_y+\hat{k}_y}+\omega)^2+\delta^2}
\right],
\nonumber
\\
& &
\label{eqn:pi_00}
\\
{\rm Im}\pi_{\rm 1-loop}(k_y,\omega)
&=&
-\int_{\frac{\pi}{2}-\hat{k}_y}^{\frac{\pi}{2}-\frac{\hat{k}_y}{2}}
\!\!
d p_y
\left[
\delta(\epsilon_{p_y}-\epsilon_{p_y+\hat{k}_y}-\omega)
+
\delta(\epsilon_{p_y}-\epsilon_{p_y+\hat{k}_y}+\omega)
\right]. 
\end{eqnarray}
The real part is given by the principle integral. 
The delta function of the imaginary part means the energy conservation for the 
particle-hole excitation. 
The $k_x$-independence is the result of the anisotropy of the striped Hall 
gas. 
The numerical results of $\pi_{\rm 1-loop}(k_y,\omega)$ are given in 
Fig.~\ref{Pi}. 
The $\omega$ region in Fig.~\ref{Pi} 
where ${\rm Im}\pi_{\rm 1-loop}(k_y,\omega)$ 
takes finite value corresponds to the particle-hole excitation region. 
At the lower boundary of the particle-hole excitation region, 
${\rm Im}\pi_{\rm 1-loop}(k_y,\omega)$ approaches zero. 
At the upper boundary of the particle-hole excitation region, 
${\rm Im}\pi_{\rm 1-loop}(k_y,\omega)$ becomes infinite negatively.  
In contrast to the ordinary 2D electron gas system in which the Fermi 
surface is sphere shape, 
the particle-hole excitation region has a gap for a finite $k_y$ due to 
one-dimensional nature of the striped Hall gas or 
the strip shape of the Fermi surface. 

\subsection{Random phase approximation (bubble and ladder diagram)}
\begin{figure}[h]
 \centerline{\includegraphics[width=.7\linewidth]{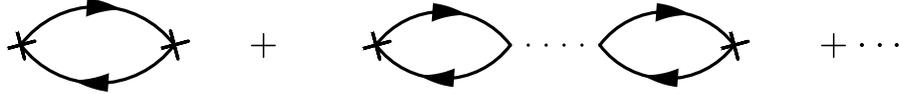}}
 \caption{Bubble diagram summation (RPA)}
 \label{fig:bubble}
\end{figure}

\begin{figure}
 \centerline{\includegraphics[width=.7\linewidth]{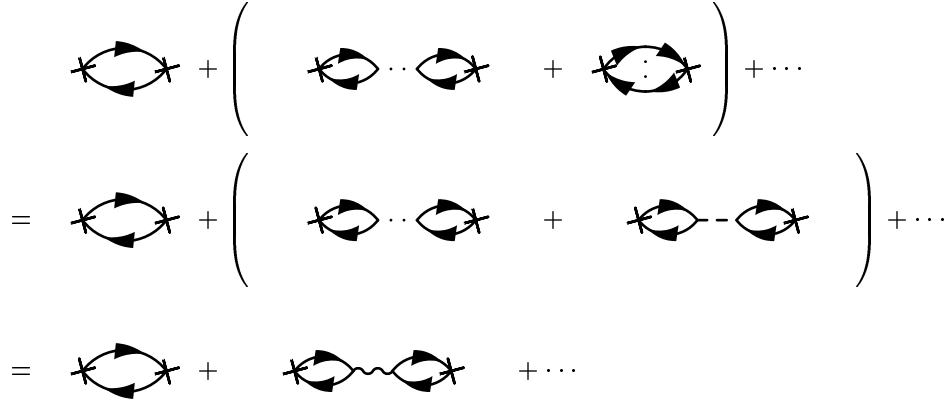}}
 \caption{Bubble + Ladder diagram summation (GRPA).
The dotted and dashed line means $W$ and $-\tilde{W}$, respectively. 
The wavy line is $W_{\rm eff}$. }
 \label{fig:grpa}
\end{figure}

\hspace*{1em}
The quantum fluctuation beyond the one-loop order is calculated in the RPA 
which is the summation of the geometric series of the one-loop
polarization function as shown in Fig.~\ref{fig:bubble}. 
Under a homogeneous magnetic field, 
we can also sum up bubble and ladder diagrams as shown in
Fig.~\ref{fig:grpa}, which is called the generalized RPA 
(GRPA).\cite{Cote Mac} 
As an example, the two-loop diagram is calculated in Appendix~\ref{2-loop}. 


The RPA polarization function is written as 
\begin{eqnarray}
\pi_ {\rm RPA}({\bf k},\omega)
&=&
\pi_{00}^{(0)}({\bf k},\omega)
+
\sum_n
\pi_{0n}^{(0)}({\bf k},\omega) 
W_n({\bf k}) \pi_{n0}^{(0)}({\bf k},\omega)
\nonumber
\\
& &
\qquad 
+
\sum_{nm}
\pi_{0n}^{(0)}({\bf k},\omega) 
W_n({\bf k})
\pi_{nm}^{(0)}({\bf k},\omega)
W_m({\bf k})
\pi_{m0}^{(0)}({\bf k},\omega)
+
\cdots
\nonumber 
\\
&=&
\sum_n 
\pi_{0n}^{(0)}({\bf k},\omega) 
\bigl[1- W(k) \pi^{(0)}({\bf k},\omega) \bigr]^{-1}_{n0}, 
\label{eq:bubble}
\end{eqnarray}
where we define
\begin{eqnarray}
\pi_{nm}^{(0)}({\bf k},\omega)
\equiv
-i
\int_{-\infty}^{\infty} \!\! \frac{dp_0}{2\pi} 
\int_{\rm MBZ} \!\! \frac{d^2p}{(2\pi)^2} \;\;
\tilde{G}^{(0)}_{{\bf p},p_0}
\;
\tilde{G}^{(0)}_{{\bf p}+\hat{\bf k},p_0+\omega}
\;
e^{i{\bf p}\times({\bf n}-{\bf m})},
\label{eq:pi_nm}
\end{eqnarray}
$W_n({\bf k}) \equiv V_l({\bf k}+2\pi \tilde{\bf n})$, and 
${\bf a}\times{\bf b}\equiv a_xb_y-a_yb_x$. 
Here $\tilde{\bf n}=(n_x/r_s,r_s n_y)$. 
Since $\tilde{G}^{(0)}_{{\bf p},p_0}$ is independent of $p_x$ for the 
striped Hall gas, we can integrate Eq.~(\ref{eq:pi_nm}) over $p_x$, 
and then
\begin{eqnarray}
\pi_{nm}^{(0)}({k}_y,\omega)
=
-i
\int_{-\infty}^{\infty} \!\! \frac{dp_0}{2\pi} 
\int_{\rm MBZ} \!\! \frac{dp_y}{2\pi} \;\;
\tilde{G}^{(0)}_{p_y,p_0}
\;
\tilde{G}^{(0)}_{p_y+\hat{k}_y,p_0+\omega}
\;
e^{-ip_y(n_x-m_x)}
\delta_{n_y-m_y,0}.
\label{eq:pi_nm_stripe}
\end{eqnarray}
$\pi^{(0)}_{nn}$ is independent of $n$ and equivalent to 
the one-loop polarization function 
$\pi_{\rm 1-loop}(k_y,\omega)$. 
Here, 
$\sum_n=\sum_{n=-\infty}^{+\infty}$ appears as the result of dividing
the infinite momentum integral region by the MBZ.
$\bigl[1- W(k) \pi^{(0)}({\bf k},\omega) \bigr]^{-1}_{n0}$ means the
$(n0)$-element of the inverse matrix of  
$\bigl[1- W(k) \pi^{(0)}({\bf k},\omega) \bigr]^{-1}$. 
In the (G)RPA, 
the fluctuation for the transverse to the stripe are included 
through the argument of exp in Eq.~(\ref{eq:pi_nm_stripe}). 
When the integer $n_x$ increases, 
the effects of the Coulomb interaction between different stripes increase because
$p_y$ direction corresponds to the $x$ direction. 

A peculiar property of a 2D system in a magnetic field is that the ladder 
diagram take a similar form with the bubble diagram in one LL.\cite{Cote Mac} 
In the presence of a magnetic field, there exists a duality relation between 
the direct term and the exchange term (see Appendix~\ref{2-loop}). 
Owing to this property, 
bubble and ladder diagrams are able to be summed up to the infinite order. 
The intra-LL polarization function in the GRPA is written as
\begin{eqnarray}
\pi_ {\rm GRPA}({\bf k},\omega)
&=&
\pi_{00}^{(0)}({\bf k},\omega)
+
\sum_n
\pi_{0n}^{(0)}({\bf k},\omega)
 \{W_n(k)-\tilde{W}_n(k)\} \pi_{n0}^{(0)}({\bf k},\omega)
\nonumber
\\
& &
\qquad
+
\sum_{nm}
\pi_{0n}^{(0)}({\bf k},\omega)
\{ W_n(k)-\tilde{W}_n(k) \}
\pi_{nm}^{(0)}({\bf k},\omega)
\{W_m(k)-\tilde{W}_m(k)\} \pi_{m0}^{(0)}({\bf k},\omega)
+
\cdots
\nonumber
\\
&=&
\sum_n 
\pi_{0n}^{(0)}({\bf k},\omega)
\bigl[1- W^{\rm eff}(k) \pi^{(0)}({\bf k},\omega) \bigr]^{-1}_{n0}, 
\label{eq:RPA_ge}
\end{eqnarray}
where we define $W_n^{\rm eff}(k)\equiv W_n(k)-\tilde{W}_n(k)$. 
Here, 
$\tilde{W}_n(k)\equiv 
\tilde{V}_l(\frac{k_y}{2\pi}+\tilde{n}_y,\frac{k_x}{2\pi}+\tilde{n}_x)$, 
and 
$\tilde{V}_l({\vec \alpha})=\int \frac{d^2\beta}{(2\pi)^2}V_l(\beta)
e^{i{\vec \beta}\cdot{\vec \alpha}}$. 
The negative sign in front of $\tilde{W}_n(k)$ is due to the exchange between 
the electron and hole in the ladder diagram. 

\begin{figure}[t]
\centerline{
 \includegraphics[width=.7\linewidth]{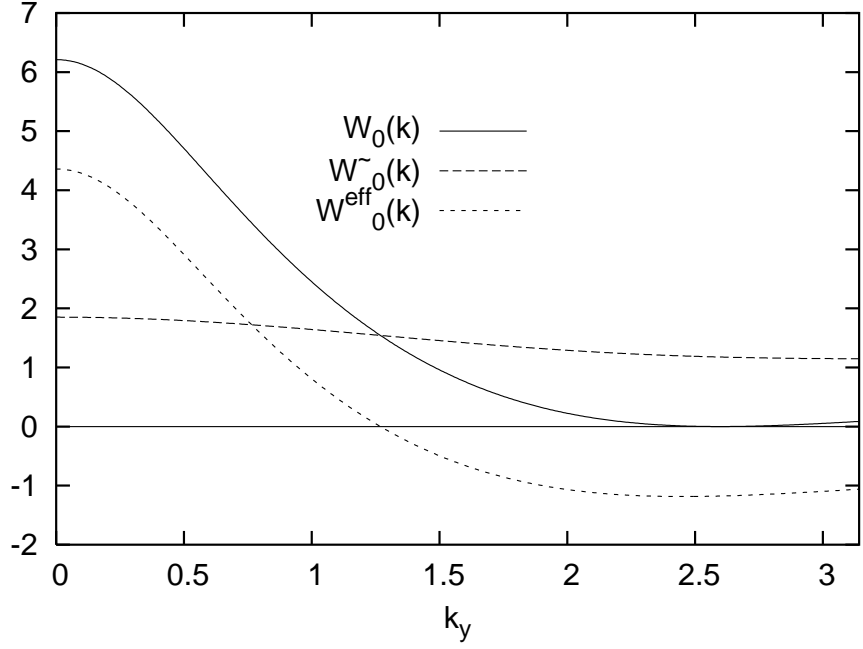}
}
\caption{Interaction $\tilde{W}_0(k)$, $W_0(k)$ and
 $W_0^{\rm eff}(k)$ at $l=2$ LL. 
We set $k_x=\pi/4$. The units of interactions and $k_y$ are 
$q^2/a$ and $1/a$, respectively. 
}
\label{fig:potential}
\end{figure}

The effective interaction $W_0^{\rm eff}(k)$ is shown in
Fig.~\ref{fig:potential} at the $l=2$ LL. 
The effective interaction $W_0^{\rm eff}(k)$ is positive for small $k$ 
and negative for large $k$. 
Thus the bubble diagram is dominant in small $k$ region and 
ladder diagram is dominant in large $k$ region. 
As shown later, the negative contribution of the effective interaction in the 
ladder diagram changes the behavior of the dielectric function and 
plasmon drastically. 

\hspace*{1em}
\subsection{Dielectric function}
We are able to obtain useful physical information of the system 
through the dielectric function. 
The RPA dielectric function is the denominator of the polarization
function in the RPA. 
In the RPA, the pole of the polarization function, 
which is equivalent to the zero of the RPA dielectric constant, 
provides a spectrum of the plasmon excitation. 

\begin{figure}[h]
\centerline{
\includegraphics[width=.7\linewidth]{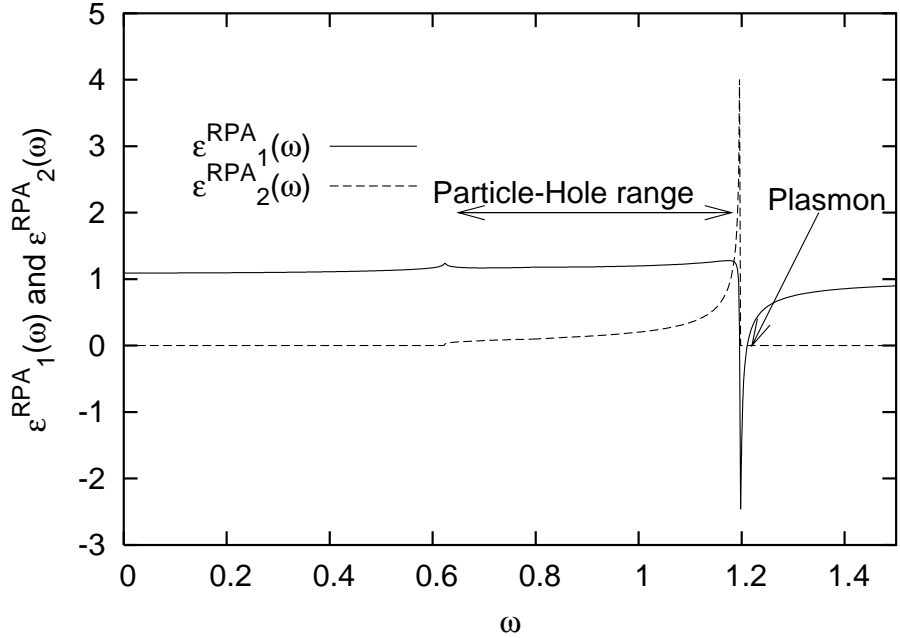}
}
\caption{$\omega$-dependence of $\epsilon^{\rm RPA}$ which includes only 
bubble diagrams for $(k_x,k_y)=(\frac{\pi}{4 r_s a},\frac{1.5 r_s}{a})$. 
The unit of $\omega$ is $q^2/(a\hbar)$. 
}
\label{fig:die_b}
\end{figure}

In the RPA, 
the dielectric function is defined by 
\begin{eqnarray} 
\epsilon^{\rm{RPA}}({\bf k},\omega)
=
1-W_0({\bf k}) \pi_{\rm 1-loop}(k_y,\omega),
\end{eqnarray}
and in the GRPA, it is defined by 
\begin{eqnarray} 
\epsilon^{\rm{GRPA}}({\bf k},\omega)
=
1-
\frac{W_0({\bf k}) 
\pi_{\rm 1-loop}(k_y,\omega)}
{
1+\tilde{W}_0 ({\bf k})
\pi_{\rm 1-loop}(k_y,\omega)
}
. 
\end{eqnarray}
Here, 
we consider only $n=0$ term in Eq.~(\ref{eq:RPA_ge}) for the first
approximation. 
The numerical results of the dielectric function at the RPA 
are shown in Fig.~\ref{fig:die_b} at 
typical $\mathbf{k}$-value $(k_x,k_y)=(\pi/4,1.5)$. 
The imaginary part $\epsilon^{\rm RPA}_2$ 
of $\epsilon^{\rm RPA}({\bf k},\omega)$ 
always takes a positive value, 
and the real part $\epsilon^{\rm RPA}_1$ 
of $\epsilon^{\rm RPA}({\bf k},\omega)$ 
has two zeros. 
$\epsilon^{\rm RPA}({\bf k},\omega)$ is finite at the particle-hole excitation
range. 

\begin{figure}[h]
\centerline{
\includegraphics[width=.7\linewidth]{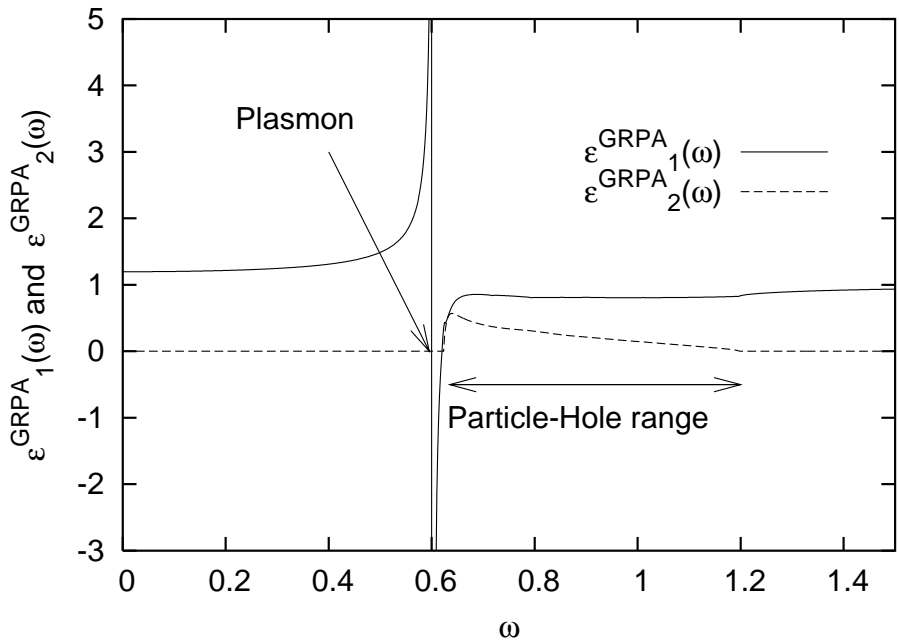}
}
\caption{$\omega$-dependence of $\epsilon^{\rm GRPA}$ which includes 
both bubble and ladder diagrams for $(\frac{\pi}{4 r_s a},\frac{1.5 r_s}{a})$. 
The unit of $\omega$ is $q^2/(a\hbar)$. 
}
\label{fig:die_l}
\end{figure}

In the GRPA, the numerical results of $\epsilon^{\rm GRPA}$ 
are shown in Fig.~\ref{fig:die_l}. 
As shown in Fig.~\ref{fig:potential}, 
the effective potential $W_0^{\rm eff}(k)$ has a negative region in contrast
to $W_0(k)$, 
which causes the drastic change of the dielectric function 
behavior. 
Thus, the point where both $\epsilon^{\rm RPA}_1$ 
and $\epsilon^{\rm RPA}_2$ are zero, 
which corresponds to the pole of the plasmon, moves over the particle-hole 
range in GRPA (compare Fig.~\ref{fig:die_b} with Fig.~\ref{fig:die_l}).

\subsection{Plasmon}
The pole of the (G)RPA polarization function gives the excitation mode
associated with the charge fluctuation, that is plasmon. 
The pole of the (G)RPA polarization function $\pi_{\rm (G)RPA}$ 
is zero of $\epsilon^{\rm (G)RPA}({\bf k},\omega)$. 
The plasmon appears at the outside range of the particle-hole pair 
regime where $\epsilon^{\rm (G)RPA}_2({\bf k},\omega)$ takes finite values. 

\begin{figure}[t]
\centerline{
\includegraphics[width=.7\linewidth]{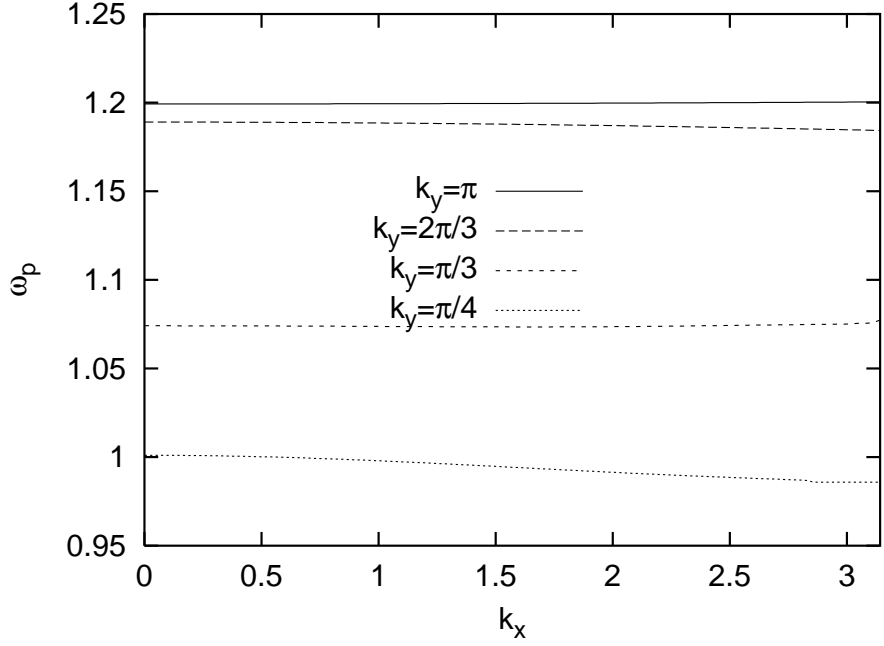}
}
\caption{$k_x$-dependence of plasma frequency
 $\omega_p(k_x,k_y)$ for the RPA in the $n=0$ case.  
The unit of $\omega_p(k_x,k_y)$ and $k_x$ is $q^2/(a\hbar)$ and $1/a$, 
respectively. }
\label{fig:plasma_b}
\end{figure}

\begin{figure}[h]
\centerline{
\includegraphics[width=.7\linewidth]{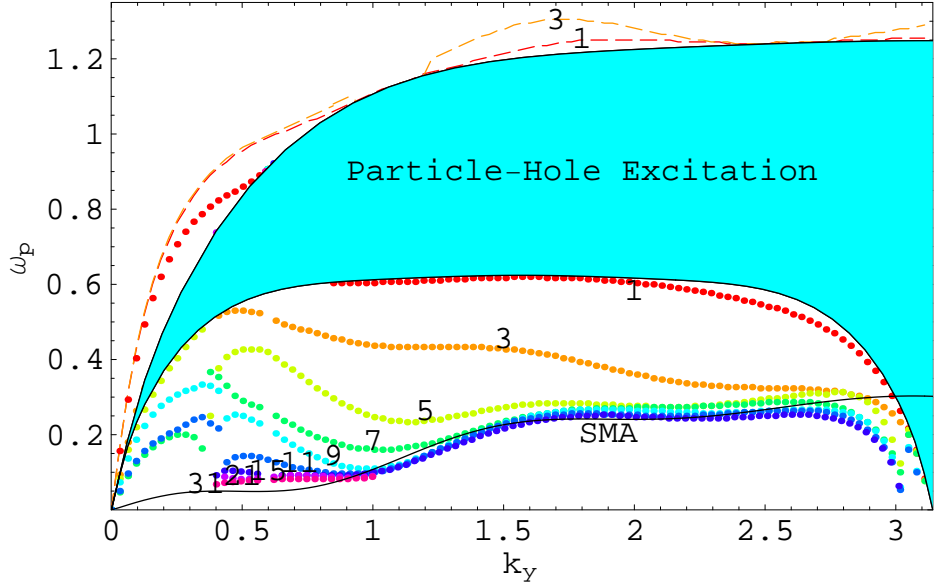}
}
\caption{$k_y$-dependence of the plasma frequency
 $\omega_p(k_x=\pi/4,k_y)$ for the RPA and the GRPA
with the particle-hole excitation region and the phonon frequency. 
The number means the dimension of the matrix 
$\left[1- W^{\rm eff}(k) \pi^{(0)}({\bf k},\omega) \right]^{-1}$. 
The dashed line is the plasmon in the RPA, and the dotted line is the one in
 the GRPA. 
The solid line is the phonon derived in the SMA.\cite{SMA} 
The behavior of $\omega_p$ approaches to the phonon one in the SMA when the
 matrix dimension increases. 
The particle-hole excitation region is unchanged and remains 
in the (G)RPA.
The unit of $\omega_p(k_x,k_y)$, $k_x$ and $k_y$ is 
$q^2/(a\hbar)$, $1/(r_s a)$ and $r_s/a$, 
respectively. 
}
\label{fig:plasma_l}
\end{figure}

First, we see the case of considering only $n=0$ term of 
Eq.~(\ref{eq:bubble}) and Eq.~(\ref{eq:RPA_ge}). 
The plasma frequency given by solving the pole of $\pi_{\rm
(G)RPA}$ 
is shown in Fig.~\ref{fig:plasma_b} and Fig.~\ref{fig:plasma_l}. 
At $k_x\neq 0$, the plasma frequency always approaches zero 
as $k_y\rightarrow 0$. 
On the other hand, at $k_y\neq 0$ the plasma frequency remains a finite
value at $k_x=0$. 
The difference of the plasmon behavior between $k_x$ and $k_y$ direction is 
the result of the spontaneous breaking of the magnetic translation and 
rotation symmetry of the striped Hall gas.
For the long wavelength limit and 
$\epsilon_{p_y+k_y}-\epsilon_{p_y}\ll \omega_p$, 
the plasma frequency $\omega_p$ 
rises like $\vert k_y\vert
\sqrt{-\ln\vert k_y\vert}/{|{\bf k}|^{1/2}}$ for taking only $n=0$ term. 
The origin of the square root behavior is the Coulomb interaction $1/r$ 
in two dimensional space. 
The logarithmic correction is caused by the divergent Fermi
velocity due to the Coulomb interaction. 

For $n=0$ case, in the GRPA, 
the $k_y$ dependence of the plasma frequency separates two region.
The one region is lower than the particle-hole pair excitation region,
where $W_0^{\rm eff}(k)$ is negative. 
The another region is higher than the particle-hole pair excitation region,
where $W_0^{\rm eff}(k)$ is positive. 
At the positive interaction region where bubble diagrams are dominant, 
the plasma frequency appears above the particle-hole excitation 
and collapses into the particle-hole excitation. 
On the other hand, at the negative interaction region 
where ladder diagrams are dominant, 
it appears below the particle-hole excitation. 
This negative interaction dominant state is considered as a low energy 
bound state due to the effective attractive interaction $W_0^{\rm eff}(k)$. 
At the long wave length range, the plasma frequency behaves as the same as
the case of $n=0$ RPA. 

Next we include finite $n$ terms of
Eq.~(\ref{eq:bubble}) and Eq.~(\ref{eq:RPA_ge}). 
In the RPA, the plasma frequency is slightly larger than $n=0$ case at 
large $k_y$ region (see Fig.~\ref{fig:plasma_l}). 
On the other hand, in the GRPA, the plasma frequency is smaller than $n=0$ 
case for wide $k_y$ region (see Fig.~\ref{fig:plasma_l}), 
and approaches to the phonon frequency associated with the density 
fluctuation derived in the SMA.\cite{SMA} 
In the striped Hall gas, the phonon is a Nambu-Goldstone mode associated 
with the spontaneous symmetry breaking for the magnetic translation 
due to $Q_Y$ in Eq.~(\ref{eq:Q}).\cite{SMA} 
Increasing $n$ in Eq.~(\ref{eq:RPA_ge}) means that the polarization
function includes effects of $n$th MBZ increasingly. 
For the large $n$ value, 
the argument of $W_n^{\rm eff}(k)$ is in the large $k$ range as seen in 
Fig.~\ref{fig:potential}. 
Since the charge density is the same as the electron density, 
it is reasonable that the plasmon associated with the charge fluctuation 
is the same as the phonon associated with the density fluctuation. 
For the small $k_y$ range, the convergence of the numerical calculation 
for large $n$ is not good because of the singular behavior of the 
polarization function near $k_y=0$. 

\subsection{Correlation energy}
\begin{figure}
\centerline{\includegraphics[width=.7\linewidth]{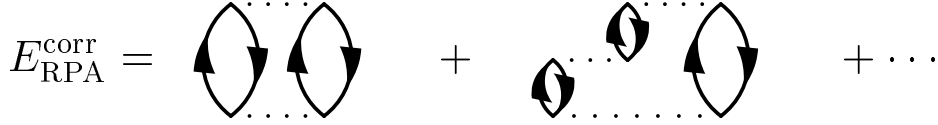}}
\caption{The chain diagram for RPA correlation energy.}
\label{fig:bubble_corre} 
\end{figure}

\hspace*{1em}
We calculate the (G)RPA correlation energy in this section. 
As is well known, the correlation energy is given by 
the virtual coupling constant $\lambda$ integration\cite{fetter} 
\begin{eqnarray}
E^{\rm total}
=
E^{\rm HF}
+\int_0^1 d\lambda \frac{1}{2}
\int_{-\infty}^{\infty}\frac{d^2 k}{(2\pi)^2}
\tilde{V}({\bf k})
\langle E(\lambda)|
\;_\circ^\circ \;
\tilde{\rho}({\bf k}) \tilde{\rho}(-{\bf k})
\;_\circ^\circ \;|E(\lambda)\rangle,
\label{Eq:cene}
\end{eqnarray}
where $\tilde{\rho}({\bf k})=j^0({\bf k})$, 
$\tilde{V}({\bf k})=\frac{2\pi q^2}{|{\bf k}|}$ 
is the Fourier transformed Coulomb interaction, and 
$\vert E(\lambda)\rangle$ is the ground state for the system with the 
virtual coupling constant $\lambda$. 
The second term of Eq.~(\ref{Eq:cene}) is the correlation energy. 
By replacing $\tilde{\rho}({\bf k})$ with 
$\tilde{\rho_\ast}({\bf k})$ and 
$\tilde{V}({\bf k})$ with $V_l({\bf k})$ in Eq.~(\ref{Eq:cene}), 
the LL projected correlation energy is represented by a vNL basis: 
\begin{eqnarray}
E^{\rm corr}
&=&
\frac{1}{2}\int_0^1 d\lambda
\int_{-\infty}^{\infty}\frac{d^2 k}{(2\pi)^2}
V_l({\bf k})
\langle E(\lambda)|\;_\circ^\circ \;
\tilde{\rho_\ast}({\bf k}) \tilde{\rho_\ast}(-{\bf k})
\;_\circ^\circ \;|E(\lambda)\rangle 
\nonumber
\\
&=&
\frac{1}{2}\int_0^1 d\lambda
\int_{\rm MBZ}
 \frac{d^2\hat{k}}{(2\pi)^2}
\frac{d^2 p_1}{(2\pi)^2}\frac{d^2 p_2}{(2\pi)^2}
\sum_{n}V_l ({\bf k}+2\pi \tilde{{\bf n}})
e^{-\frac{i}{2\pi}\hat{k}_x(p_1-p_2)_y
+i({\bf p}_1-{\bf p}_2)\times {\bf n}}
\nonumber
\\ 
& &
\qquad
\times
\langle E(\lambda)|\;_\circ^\circ
b_{l,{\bf p}_1+\hat{\bf k}}^\dagger b_{l,{\bf p}_1}
b_{l,{\bf p}_2}^\dagger b_{l,{\bf p}_2+\hat{\bf k}}\;_\circ^\circ
|E(\lambda)\rangle.
\label{Eq:rpaene}
\end{eqnarray}
The integer $n$ is caused by dividing the $k$-integral region
into summation of one MBZ. 
The RPA correlation energy 
is the sum of the chain diagram of bubble as shown in 
Fig.~\ref{fig:bubble_corre} 
which is derived by the perturbative calculation about the virtual coupling
constant $\lambda$. 
The density correlation function part in Eq.~(\ref{Eq:rpaene}) 
is calculated as
\begin{eqnarray}
& &
\int_{\rm MBZ}
\frac{d^2 p_1}{(2\pi)^2}\frac{d^2 p_2}{(2\pi)^2}
\sum_{n}V_l ({\bf k}+2\pi \tilde{{\bf n}})
e^{-\frac{i}{2\pi}\hat{k}_x(p_1-p_2)_y
+i({\bf p}_1-{\bf p}_2)\times {\bf n}}
\nonumber
\\ 
& &
\qquad
\times
\langle E(\lambda)|\;_\circ^\circ
b_{l,{\bf p}_1+\hat{\bf k}}^\dagger b_{l,{\bf p}_1}
b_{l,{\bf p}_2}^\dagger b_{l,{\bf p}_2+\hat{\bf k}}\;_\circ^\circ
|E(\lambda)\rangle
\nonumber
\\
&\stackrel{\rm RPA}{\simeq}&
i \sum_{mnp}
\pi^{(0)}_{mn}(k) W_n(k) \pi^{(0)}_{np}(k) 
\lambda W_p(k)
\left(
\frac{1}{1-\pi^{(0)}(k) \lambda W(k)}
\right)_{pm}
\label{eq:rpa1}
\end{eqnarray}
Here $\pi^{(0)}_{nm}$ is given in Eq.~(\ref{eq:pi_nm}).  
We take only the diagonal matrix elements contribution and approximate 
Eq.~(\ref{eq:rpa1}) as
\begin{equation}
i\,
\sum_n ( \pi^{(0)}_{nn}(k)W_n(k) )^2
\frac{\lambda}{1-\pi^{(0)}_{nn}(k) \lambda W_n(k)}. 
\label{Eq:rpa}
\end{equation}

In the summation about $n$, 
the contribution of $n\geq 1$ terms are negligible because of the
Gaussian factor in $V_l({\bf k})$. 
So we consider only the $n=0$ term. 
By substituting Eq.~(\ref{Eq:rpa}) into Eq.~(\ref{Eq:rpaene}), 
the integral about $\lambda$ gives the RPA correlation energy 
$E_{\rm RPA}^{\rm corr}$ as 
\begin{eqnarray}
E_{\rm RPA}^{\rm corr}
=
-\frac{i }{2}
\int_{-\infty}^{\infty}
\frac{d \omega}{2\pi}
\int_{\rm MBZ}\frac{d^2\hat{k}}{(2\pi)^2}
\left\{
\log{( \epsilon^{\rm RPA}({\bf k},\omega) )} 
+ \pi_{\rm 1-loop}(k_y,\omega)
W_0(k)
\right\}. 
\label{Eq:E_c}
\end{eqnarray} 
The numerical estimates of $E^{\rm HF}$, the real part of 
$E_{\rm RPA}^{\rm corr}$ 
and the total energy
$E^{\rm total}$ per particle at the half-filled $l=2$ LL are 
obtained as follows: 
\begin{eqnarray}
E^{\rm HF}&=&-0.7706, \nonumber \\
E^{\rm corr}_{\rm RPA}&=& -0.0341,\\
E^{\rm total}&=& -0.8047, \nonumber
\end{eqnarray}
where the energy unit is $\frac{q^2}{a}$. 
The total energy is lowered by the RPA correlation energy significantly. 

\begin{figure} 
 \centerline{\includegraphics[width=.7\linewidth]{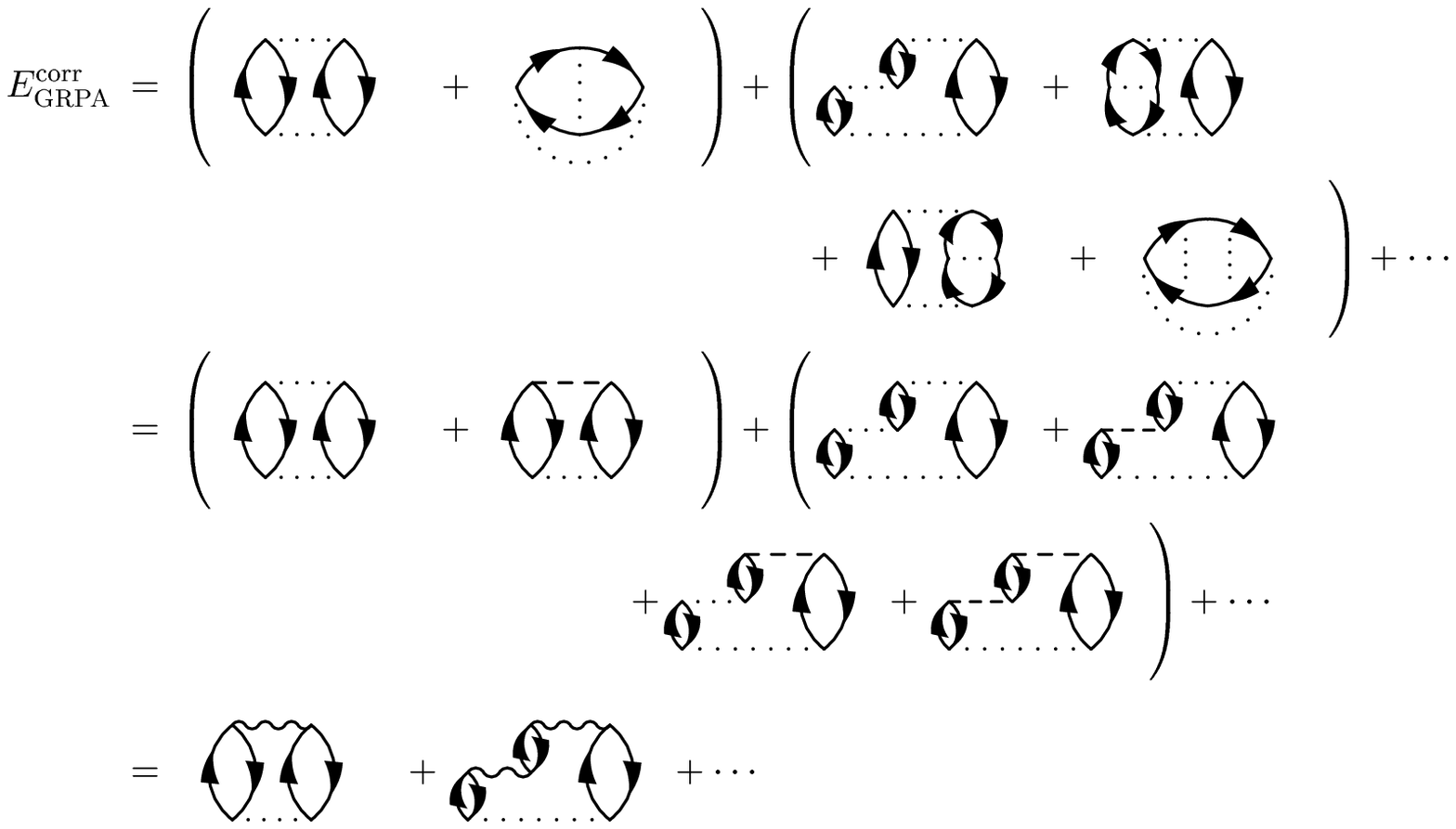}}
 \caption{The chain diagram for GRPA correlation energy.}
 \label{fig:grpa_corr}
\end{figure}

GRPA correlation energy 
is the sum of the chain diagram of bubble and ladder 
as shown in Fig.~\ref{fig:grpa_corr}.  
Corresponding to Eq.~(\ref{Eq:rpa}) in RPA, 
Eq.~(\ref{eq:rpa1}) is approximated in GRPA as 
\begin{eqnarray}
i\,
\sum_n ( \pi^{(0)}_{nn}(k))^2W_n({\bf k})W_n^{\rm eff}({\bf k})
\frac{\lambda}{1-\pi^{(0)}_{nn}(k) \lambda W^{\rm eff}_n({\bf k})}. 
\nonumber
\end{eqnarray}
Considering $n=0$ term, 
the GRPA correlation energy $E_{\rm GRPA}^{\rm corr}$ is written by 
$\pi_{\rm 1-loop}(k_y,\omega)$ as 
\begin{eqnarray}
E_{\rm GRPA}^{\rm corr}
=
-\frac{i }{2}
\int_{-\infty}^{\infty}
\frac{d \omega}{2\pi}
\int_{\rm MBZ}\frac{d^2\hat{k}}{(2\pi)^2}
\left\{ 
\left(
\frac{W_0(k)}{W^{\rm eff}_0(k)}
\right)
\log{\left[ 1-W^{\rm eff}_0(k) \pi_{\rm 1-loop}(k_y,\omega) \right]} 
+ \pi_{\rm 1-loop}(k_y,\omega) W_0(k)
\right\}. 
\label{Eq:E_c_grpa}
\end{eqnarray} 
The numerical estimates of the real part of 
$E_{\rm GRPA}^{\rm corr}$ 
and the total energy
$E^{\rm total}$ per particle at the half-filled $l=2$ LL are 
obtained as follows: 
\begin{eqnarray}
E^{\rm corr}_{\rm GRPA}&=& -0.0385,\\
E^{\rm total}&=& -0.8090.
\nonumber 
\end{eqnarray}
GRPA correlation energy is slightly lower than RPA correlation energy. 

Yoshioka calculated the corresponding HF energy for the ACDW 
state.\cite{daijiro} 
The ACDW state has anisotropic energy gaps in $p_x$ and $p_y$-direction. 
Our striped Hall gas, on the other hand, has an energy gap only in 
$p_y$-direction. 
The numerical value of Yoshioka's HF energy is $E^{\rm Y}=-0.7763$. 
Our total energy $E^{\rm total}$ of the ground state is smaller than 
$E^{\rm Y}$, 
hence our striped Hall gas including the quantum fluctuation at the 
(G)RPA level is more stable than the ACDW. 
On the other hand, 
Shibata and Yoshioka studied the ground state phase of 2D electrons in
$l=2$ LL by a density matrix renormalization group (DMRG) method, 
which is a numerical calculation of a small system improved by the 
renormalization group method.\cite{Shibata} 
The DMRG results seem to predict the striped Hall gas at the 
half-filled higher LL. 
The total energy given by the DMRG is 
$E^D=-0.796\pm0.004$ \cite{Shibata_pri} 
which is smaller than $E^{\rm Y}$, 
whereas it is close to $E^{\rm total}$. 

The one-electron energy of the ACDW state has a gap 
in both $p_y$ and $p_x$ direction, whose value is 
about 1 K.\cite{bubble_fertig} 
Hence, this state is insulator in $x$ and $y$ direction and the
gap structure causes the quantization of the Hall conductance. 
However, experiments show the huge anisotropic resistivity and the Hall 
conductance is not quantized at several mK.\cite{Lilly,Du} 
Since the striped Hall gas has the anisotropic Fermi surface, 
the $x$ direction is insulator whose gap energy is the cyclotron energy,
and the $y$ direction is metal in which the electron gas state 
realizes. 
Therefore the striped Hall gas is more consistent with experiments 
than ACDW. 
Moreover the comparison of the correlation energy with the ACDW and 
results of DMRG seems to support our striped Hall gas. 

\section{Summary}
\label{IV}
Using the one-loop polarization function which includes only intra-LL effects, 
the dielectric function, the plasma frequency, and correlation energy 
are calculated in the (G)RPA for the striped Hall gas. 
The characteristic feature of the plasma frequency is anisotropic gapless 
behavior. 
The anisotropy is due to the spontaneous breaking of rotational symmetry 
and the gapless feature comes from two-dimensionality of the system. 
The anisotropic plasma frequency will be observed by some experiments: 
e.g. a surface acoustic wave. 
The numerical result of GRPA plasmon suggests that the plasmon in the 
striped Hall gas is the same as phonon in the striped Hall gas 
which is the Nambu-Goldstone mode due to the the 
spontaneous breaking of translational symmetry. 
In contrast to the quantum Hall smectic\cite{Anna} derived by edge current 
picture and TDHFA applying to the ACDW, 
this excitation state reflects the striped Hall gas state. 
It is shown that the quantum fluctuation effect for the striped Hall gas 
substantially reduces the total energy in (G)RPA. 
This means that the quantum fluctuation plays a important role 
in the striped Hall gas. 
The quantum Hall gas properties strongly depends on the electron self-energy
with the anisotropic Fermi surface. 
The treatment for quantum fluctuation effects to the electron self-energy 
is beyond the scope of the present paper, and is very interesting as a future problem. 

\begin{acknowledgments}
T.~A. thanks Akinori Asahara for useful comments on numerical
 calculations. 
This work was partially 
supported by the special Grant-in-Aid for Promotion of Education and 
Science in Hokkaido University and the Grant-in-Aid for Scientific Research 
on Priority area (Dynamics of Superstrings and Field Theories) 
(Grant No.13135201), provided by Ministry of Education, Culture, Sports, 
Science, and Technology, Japan, 
and by Clark Foundation and Nukazawa Science Foundation. 
\end{acknowledgments}
\appendix
\section{Hartree-Fock approximation}
\label{HF}
In the intra-LL HFA,
we reduce the Coulomb interaction term to the kinetic term 
by using a mean field 
$U(p)\equiv \langle{\rm HF}\vert b_l^\dagger({\bf p})b_l({\bf p})
\vert{\rm HF} \rangle$ 
where $\vert\rm HF\rangle$ is a many-particle state satisfying a 
self-consistency equation. 
The interaction Hamiltonian projected into the $l$th LL, 
$$
H_1
\stackrel{\rm project}{=}
\frac{1}{2}\int_{-\infty}^{\infty} \frac{d^2 k}{(2\pi)^2} 
:
\tilde{\rho_\ast}({\bf k})
V_l(k)
\tilde{\rho_\ast}(-{\bf k})
:,
$$
is approximated by the HF Hamiltonian
\begin{equation}
H^{\rm HF}=
\frac{\nu^\ast}{N}
\int_{\rm MBZ}
\!\!
\frac{d^2p}{(2 \pi)^2}\frac{d^2p^\prime}{(2\pi)^2}
v_l^{\rm HF}\!({\bf p}-{\bf p}^\prime) U({\bf p}^\prime)
b_{l}^\dagger({\bf p})b_{l}({\bf p})
- 
\frac{\nu^\ast}{2N}
\int_{\rm MBZ}
\!\!
\frac{d^2p}{(2 \pi)^2}\frac{d^2p^\prime}{(2\pi)^2}
U({\bf p}) v_l^{\rm HF}\!({\bf p}-{\bf p}^\prime) U({\bf p}^\prime). 
\label{eq:HF}
\end{equation}
Here, we define  
\begin{eqnarray}
v_l^{\rm HF}({\bf p}-{\bf p}^\prime)
\equiv
\sum_{n}
\left\{
V_l
(2\pi \tilde{\bf n})
e^{i ({\bf p}-{\bf p}^\prime)\times {\bf n}}
-V_l
( 2\pi \tilde{{\bf n}}+\tilde{{\bf p}}^\prime-\tilde{\bf p})
\right\}.
& &
\label{eq:vHF}
\end{eqnarray}
$\nu^\ast$ is a filling factor for the highest LL, and $N$ is a number
of electrons at the highest LL. 
In this paper, $\nu^\ast$ is set to 1/2. 
The states below $l-1$ th LL are fully occupied. 
The first term in Eq.~(\ref{eq:vHF}) is the Hartree potential and the
second one is the Fock potential. 
The uniform positive background charge cancels the ${\bf n}=0$ term in
the Hartree term. 
Since the state at the $l$th LL is occupied by electrons whose energy
is below a Fermi energy $\mu_{\rm F}$, 
the mean field is written as 
$U({\bf p})
=\frac{N}{\nu^\ast}\theta(\mu_{\rm F}-\epsilon_l^{\rm HF}({\bf p}))$. 
Hence, 
the self-consistency equation of one-electron energy 
$\epsilon_l^{\rm HF}({\bf p})$ reads 
\begin{equation}
\epsilon_l^{\rm HF}({\bf p})
=
\int_{\rm MBZ}\frac{d^2p^\prime }{(2\pi)^2}
v_l^{\rm HF}({\bf p}-{\bf p}^\prime) \,
\theta(\mu_{\rm F}-\epsilon_l^{\rm HF}({\bf p}^{\prime})\,).
\label{eq:self}
\end{equation}
The one-electron energy has a periodic structure 
$\epsilon_l^{\rm HF}({\bf p})=\epsilon_l^{\rm HF}({\bf p}+2\pi{\bf n})$ 
owing to $v_l^{\rm HF}({\bf p})=v_l^{\rm HF}({\bf p}+2\pi{\bf n})$. 

\section{One-electron spectrum}
\label{appe_a}
The one-electron spectrum is given by the next explicit relation
$\epsilon_l^{\rm HF}(p_y)=\epsilon_l^{\rm H}(p_y)+\epsilon_l^{\rm
F}(p_y)$, 
where 
\begin{eqnarray}
\epsilon_l^{\rm H}(p_y)
&=&
\frac{2 q^2}{\pi}\sum_{n=0}^{\infty}
V_l\!\left[ \frac{2\pi(2n+1)}{r_s}\right]
\frac{(-1)^n \cos[(2n+1)p_y]}{(2n+1)}
\\
\epsilon_l^{\rm F}(p_y)
&=&
-r_s q^2
\sum_{n=-\infty}^{\infty}
\int_{-\frac{\pi}{2}}^{
\frac{\pi}{2}} 
\frac{dk_y}{2\pi}
\int_{-\infty}^{\infty} 
\!\!\!
\frac{dk_x}{2\pi} \;
V_l\!
\left[
\sqrt{
k_x^2+r_s^2 (k_y-p_y-2\pi n)^2
}
\right]. 
\label{eq:spectrum}
\end{eqnarray}
\section{Feynman rule}
\label{Feynman_rule}
In the following calculation, 
the interaction picture is applied in perturbation theory. 
In the Heisenberg picture, 
the time-dependence of an operator $\mathcal{O}^{\rm H}(t)$ 
is defined as 
$\mathcal{O}^{\rm H}(t)=e^{iH_1 t}\mathcal{O}e^{-iH_1 t}$. 

The electron Green' function is defined by 
\begin{eqnarray}
\langle \Psi_0 | 
{\rm T} b^{\rm H}_{{\bf p}_1,t_1} b^{\dagger {\rm H}}_{{\bf p}_2,t_2}
| \Psi_0 \rangle
&\equiv&
i G({\bf p}_1,t_1-t_2)
\sum_n (2\pi)^2 
\delta^{(2)}({\bf p}_1-{\bf p}_2-2\pi {\bf n}) e^{i\phi(p_1,n)}, 
\label{eq:green_ele}
\end{eqnarray}
where $|\Psi_0 \rangle$ is the exact ground state of $H_1$, 
and $\phi(p,n)=\pi(n_x+n_y)-n_y p_x$. 
$\rm T$ means the time ordering. 
We write $b_{l}({\bf p},t)$ as $b_{{\bf p},t}$ and omit its LL index. 
The Heisenberg picture is changed to the interaction picture by using
the relation 
$b_{{\bf k},t}^{\rm H}=S(0,t)b_{{\bf k},t}S(t,0)$, 
where $S(t,t^\prime)=e^{iH^{\rm HF}t} e^{-iH_1 t}
e^{iH^{\rm HF}t^\prime} e^{-iH_1 t^\prime}$. 
Using $S(t,t^\prime)$ in Eq.~(\ref{eq:green_ele}), 
the Green function is written as a familiar form: 
\begin{eqnarray}
G({\bf k},t-t^\prime)
=
-i \sum_{n=0}^{\infty} 
\frac{(-i)^n}{n!} \int_{-\infty}^{\infty}dt_1 
\cdots \int_{-\infty}^{\infty}dt_n 
\frac
{\langle {\rm HF} |T b_{{\bf k},t} b^\dagger_{ {\bf k},t^\prime } 
V(t_1)\cdots V(t_n) 
|{\rm HF} \rangle}
{\langle {\rm HF} | S(+\infty,-\infty) |{\rm HF} \rangle}, 
\label{eq:Green_full}
\end{eqnarray}
where the interaction $V(t)$ is the residual interaction in the interaction 
picture 
\begin{eqnarray}
V(t)
=
\frac{1}{2}\int_{-\infty}^{\infty} 
\!\!\!
\frac{d^2 k}{(2\pi)^2} 
\;_\circ^\circ \;
\tilde{\rho_\ast}({\bf k},t)
V_l(k)
\tilde{\rho_\ast}(-{\bf k},t)
\;_\circ^\circ \;. 
\label{eq:v}
\end{eqnarray}
The lowest order Green function is obtained as
\begin{eqnarray}
G^{(0)}({\bf k},t-t^\prime)
&=&
-i
\left\{
\theta(t-t^\prime) \theta(\epsilon_{\bf k}-\mu_{\rm F}) 
e^{-i(t-t^\prime)(\epsilon_{\bf k}-\mu_{\rm F})} 
-
\theta(t^\prime-t) \theta(\mu_{\rm F}-\epsilon_{\bf k}) 
e^{-i(t-t^\prime)(\epsilon_{\bf k}-\mu_{\rm F})} 
\right\}.
\nonumber
\\
& &
\end{eqnarray}
Considering only intra LL effects,
we omit the LL index and take the short notation of 
$\epsilon_l^{\rm HF}({\bf k})$ as $\epsilon_{\bf k}$. 
The Fourier transformation of the free propagator reads 
\begin{eqnarray}
\tilde{G}^{(0)}_{{\bf k},\omega}
&=&
\int_{-\infty}^{\infty} dt \; e^{i\omega t} G^{(0)}({\bf k},t) 
\nonumber \\
&=&
\frac{\theta(\epsilon_{\bf k}-\mu_{\rm F})}{\omega-\epsilon_{\bf
k}+\mu_{\rm F}+i\delta}
+
\frac{\theta(\mu_{\rm F}-\epsilon_{\bf k})}{\omega-\epsilon_{\bf
k}+\mu_{\rm F}-i\delta}. 
\label{eq:free_proper}
\end{eqnarray}

In the vNL formalism, 
it is transparent way to represent the infinite ${\bf k}$-integral of
Eq.~(\ref{eq:v}) as the summation of one fundamental MBZ. 
So the residual interaction is written as 
\begin{equation}
V(t)=
\int_{\rm MBZ}\!\!
\frac{d^2\hat{k}}{(2\pi)^2} \frac{d^2 p_1}{(2\pi)^2} \frac{d^2 p_2}{(2\pi)^2} 
e^{-\frac{i}{2\pi} \hat{k}_x p_{1y}}
V_l(p_1,k,p_2) 
e^{\frac{i}{2\pi} \hat{k}_x p_{2y}}
\>\;_\circ^\circ \>
b_{t,{\bf p}_1+ \hat{\bf k}}^\dagger b_{t,{\bf p}_1 }
b_{t,{\bf p}_2}^\dagger b_{t,{\bf p}_2+ \hat{\bf k}}
\>\;_\circ^\circ \>, 
\label{eq:vertex0}
\end{equation}
where we define
\begin{eqnarray}
V_l(p_1,k,p_2)
&\equiv&
 \sum_{n=-\infty}^\infty V_l(k+2\pi \tilde{n})
e^{i({\bf p}_1-{\bf p}_2)\times {\bf n}}.
\label{eq:vpkp}
\end{eqnarray}
In stead of the infinite $k$-integral region, 
the infinite summation appears. 
The local interaction $V_l(k)$ in momentum space 
is replaced by the non-local one $V_l(p_1,k,p_2)$ 
including the phase factor in the density operator. 
For perturbative calculations, 
we use the Wick's theorem and obtain an $n$-point correlation function
for the interaction $V(t)$. 

In the following we present the Feynman diagram rule in the momentum space 
for the perturbative residual 
interaction Eq.~(\ref{eq:vertex0}). 
The phase factor due to the magnetic field makes 
the vertex and the momentum conservation factors complicate. 

(i) Draw a Feynman diagram. 
For an electron propagator, 
introduce the Green's function
\begin{eqnarray}
 \tilde{G}^{(0)}_{{\bf p},\omega}
=
\begin{minipage}[h]{.4\linewidth}
 \centerline{\includegraphics[width=.5\linewidth]{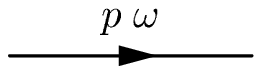}}
\end{minipage}
. 
\end{eqnarray}

(ii) For one Coulomb line which is the interaction vertex, 
a local momentum interaction is not assigned in the usual way. 
We add a non-local interaction including four electron momentum ${\bf p}_i$
with phase factor as 
\begin{eqnarray}
V_l(p_2,k,p_3) e^{-\frac{i}{2\pi}\hat{k}_x(p_1-p_4)_y}
=
\begin{minipage}[h]{.4\linewidth}
 \centerline{\includegraphics[width=.5\linewidth]{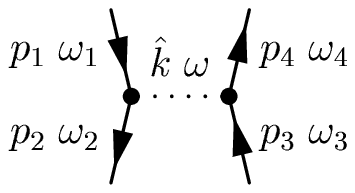}}
\end{minipage}
. 
\end{eqnarray}

(iii) Add a momentum conservation factor 
for each interaction vertices with four electron momentum ${\bf p}_i$ and 
frequency $\omega_i$
\begin{multline}
\sum_{nm}
 (2\pi)^2 
 \delta({\bf p}_1-{\bf p}_2-\hat{\bf k}-2\pi{\bf n})
 e^{i\phi(p_2,n)}
 \;
 (2\pi)^2 
 \delta({\bf p}_3-{\bf p}_4+\hat{\bf k}-2\pi{\bf m})
 e^{i\phi(p_3,m)}
\\
\times
(2\pi)^2 
\delta(\omega_1-\omega_2-\omega) \delta(\omega_3-\omega_4+\omega).
\end{multline}
The phase factor $\phi(p_i,n)$ is added to the delta function.

(iv) For one current vertex, 
a local momentum factor is also not assigned as the interaction vertex
case. 
We add a non-local factor 
\begin{eqnarray}
g_\mu(k)e^{ 
-\frac{i}{2\pi} \hat{k}_xp_{1y}+\frac{i}{4\pi}k_xk_y
}=
\begin{minipage}[h]{.4\linewidth} 
 \centerline{\includegraphics[width=.5\linewidth]{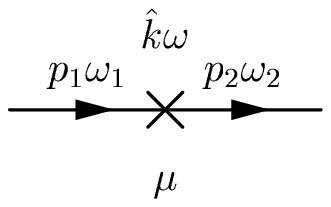}}
\end{minipage}
. 
\end{eqnarray}

(v) Add a momentum conservation factor 
for each current vertices with two electron momentum ${\bf p}_i$ and 
frequency $\omega_i$
\begin{eqnarray}
\sum_{n}
 (2\pi)^2 
 \delta({\bf p}_1-{\bf p}_2-\hat{\bf k}-2\pi{\bf n})
 e^{i\phi(p_2,n)}
(2\pi)\delta(\omega+\omega_1-\omega_2).
\end{eqnarray}
The phase factor $\phi(p_i,n)$ is added to the delta function.

(vi) Perform integral for internal momentum and add the numerical factor
\begin{eqnarray}
\left\{\frac{1}{(2\pi)^3}\right\}^I, 
\end{eqnarray}
where $I$ is the number of internal line. 
Count the electron loop number and add the factor $(-1)^L$
for $L$ electron loops. 

\section{Duality between the direct term and the exchange term}
\label{2-loop}
One of the surprising properties of the 2D electron system in a 
magnetic field is that 
the ladder diagrams take a similar form with the bubble diagram and 
the bubble and ladder diagrams can be summed up to the infinite order. 
This is caused by the duality between the direct term and the exchange term. 
In this section we show this property in the two-loop order as shown in 
Fig.~\ref{fig:2loop}. 

Following the rule of Appendix~\ref{Feynman_rule}, the left two-loop 
diagram in Fig.~\ref{fig:2loop} reads

\begin{equation}
-
\sum_n\int \frac{d\omega_1}{2\pi} \frac{d\omega_2}{2\pi}
\int_{\rm MBZ}
\frac{d^2p_1}{(2\pi)^2} \frac{d^2p_2}{(2\pi)^2} \; \;
\tilde{G}^{(0)}_{p_1-\hat{k},\omega_1-\omega}
\tilde{G}^{(0)}_{p_1,\omega_1}\tilde{G}^{(0)}_{p_2-\hat{k},
\omega_2-\omega}\tilde{G}^{(0)}_{p_2,\omega_2}
V_l(\tilde{p}_1-\tilde{p}_2+2\pi \tilde{n} )
e^{\frac{i}{2\pi}({\bf p}_1-{\bf p}_2+2\pi{\bf n})\times\hat{\bf k}} 
\label{eq:2loop}
\end{equation}
In general, the Fourier transform $\tilde{V}_l({\bf q})=\int 
\frac{d^2p}{(2\pi)^2} V_l({\bf p})e^{i{\bf p}\cdot{\bf q}}$ satisfies 
the following duality relation between the direct term and the exchange term, 
\begin{equation}
\sum_n V_l({\bf p}+2\pi{\bf n})e^{i({\bf p}+2\pi{\bf n})
\cdot{\bf q}}=\sum_n\tilde{V}_l({\bf q}+{\bf n})e^{-i{\bf p}
\cdot{\bf n}}.
\end{equation}
Using this relation, Eq.~(\ref{eq:2loop}) is written as 
\begin{eqnarray}
& &
-\sum_n
\int \frac{d\omega_1}{2\pi} \frac{d\omega_2}{2\pi}
\int_{\rm MBZ}
\frac{d^2p_1}{(2\pi)^2} \frac{d^2p_2}{(2\pi)^2} \; \;
\tilde{G}^{(0)}_{p_1-\hat{k},
\omega_1-\omega}\tilde{G}^{(0)}_{p_1,\omega_1}\tilde{G}^{(0)}_{p_2-\hat{k},
\omega_2-\omega}\tilde{G}^{(0)}_{p_2,\omega_2}
\tilde{V}_l \!\!
\left(
\frac{k_y}{2\pi}+\tilde{n}_y,
-\frac{k_x}{2\pi}-\tilde{n}_x
\right)
e^{i{\bf n}\times({\bf p}_1-\bf{p}_2)} \nonumber\\
&=&
-\sum_n\pi_{0n}(k)\tilde{W}_n(k)\pi_{n0}(k).
\end{eqnarray}
This result is equivalent to two bubbles connected with the interaction 
${\tilde W}(k)$ as the right two-loop diagram in Fig.~\ref{fig:2loop}. 
\begin{figure}[h] 
 \centerline{\includegraphics[width=.7\linewidth]{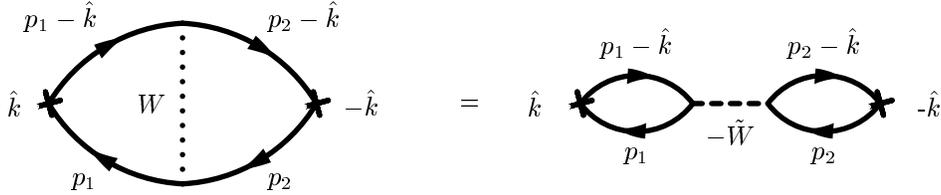}} 
 \caption{The two-loop ladder diagram and corresponding two-loop 
 bubble diagram}
 \label{fig:2loop}
\end{figure}


\end{document}